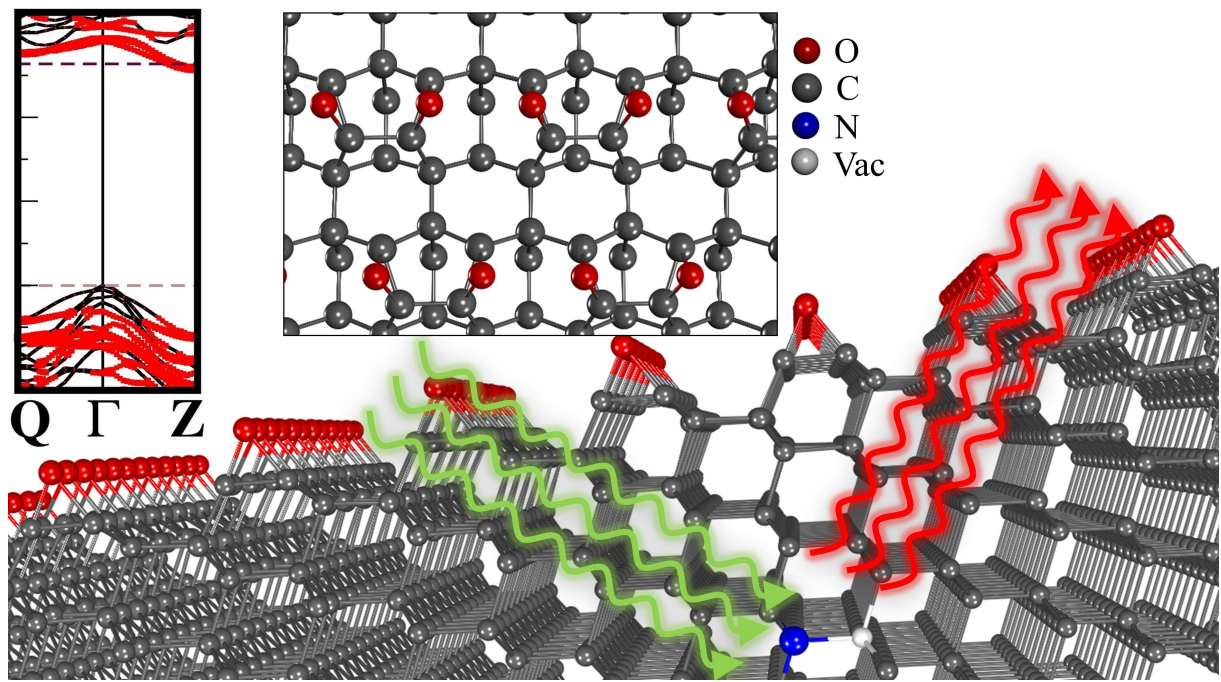

# Oxygenated (113) diamond surface for nitrogen-vacancy quantum sensors with preferential alignment and long coherence time from first principles

*Song Li[1], Jyh-Pin Chou[1,\*], Jie Wei[1], Minglei Sun[2], Alice Hu[1,\*], and Adam Gali[3,4]*

[1] Department of Mechanical Engineering, City University of Hong Kong, Kowloon Tong, Hong Kong, China

[2] School of Mechanical Engineering, Southeast University, Nanjing, Jiangsu 211189, China

[3] Institute for Solid State Physics and Optics, Wigner Research Centre for Physics, Hungarian Academy of Sciences, Budapest, POB 49, H-1525, Hungary

[4] Department of Atomic Physics, Budapest University of Technology and Economics, Budafoki út 8, H-1111, Budapest, Hungary

*Corresponding author  Email: jpchou@cityu.edu.hk and alicehu@cityu.edu.hk

**ABSTRACT**: Shallow nitrogen-vacancy (NV) center in diamond is promising in quantum sensing applications however its sensitivity has been limited by surface terminators and defects. There is an immediate quest to find suitable diamond surfaces for NV sensors. In this work, the surface terminators of (113) diamond to host shallow NV centers are studied by means of first principles calculations. Results indicate that complete oxygen termination of (113) diamond creates positive electron affinity with neither strain on the surface nor in-gap levels. This is a very surprising result as the commonly employed oxygenated (001) diamond surface is often defective due to the disorder created by the strain of ether groups at the surface that seriously undermine the coherence properties of the shallow NV centers. The special atomic configurations on (113) diamond surface are favorable for oxygen bonding, in contrast to (001) and (111) diamond surfaces. These simulations imply that oxygenated diamond (113) surface can be produced by conventional diamond chemical vapor deposition growth. Combining this with the ~73% preferential alignment of as-grown NV centers in (113) oriented diamond, oxygenated (113) diamond is presently supposed to be the most prospective host for NV quantum sensors.

# 1. INTRODUCTION

Negatively charged nitrogen-vacancy defect, i.e., the nitrogen-vacancy center (NV), in diamond has triggered a lot of interests and attention in quantum information processing, [1-5] plasmonics [6-7] and nanosensing applications. [8-12] The NV center contains one substitutional nitrogen atom adjacent to a vacancy, as shown in Figure 1a. Its negative charge state has a high-spin $S=1$ ground state with an extra electron coming from the diamond environment. The NV center has a preferential symmetry axis along <111> directions. It has an optically allowed triplet excited state with intermediate dark singlet states that play a role in the spin-selective non-radiative decay from the triplet excited state toward the triplet ground state. Its optical transient dipole moment is oriented along the symmetry axis. The robust spin-selective non-radiative decay in NV center can be harnessed to carry out optically detected magnetic resonance (ODMR) at single defect level at room temperature. [13] Through the coherent manipulation of the electron spin, NV center can be employed as nanoscale sensors for magnetic [11] and electric field detection, [9] ultrasensitive thermometer in living environment, [10] and fluorescent biomarkers. [14] These diverse utilities rely on the electron spin state of the NV center which is very sensitive to the ambient condition.

As the optical transition dipole moment of the NV center is defined by its symmetry axis along <111> directions in the diamond lattice, as a consequence, the most commonly grown (001) diamond has deficient photon collection efficiency. [15-16] Highly perfect NV center alignment on (111)-grown chemical vapor deposited (CVD) diamond layers have been demonstrated with significantly improved photon collection efficiency. [17-19] Recently, high-quality (111) diamond surfaces were sufficiently grown under optimized pressure and temperature conditions. [20] The main issue for (111) diamond surface is the significant uptake of residual or intentional impurities. [21-22] Moreover, strains developed on (111) layer which provide ODMR line broadening and low $T_2$ time further impose restrictions on (111) diamond for NV center application, [23-24] thus an alternative solution would substantially accelerate the development of NV sensors.

Recent reports showed that (113) diamond surface could be a potential substrate to produce NV center. [25-29] The (113) surface can be successfully synthesized through laser cutting and polishing a standard (001) surface with a 29.5° angle towards <111> direction. Previous theoretical calculation confirms the 2×1 reconstructed (113) surface is more energetically favorable than (001) 2×1 surface. [30] Further experiment indicates that similar to silicon (113) surface [31-32] the stability of the (113)-grown CVD diamond can be possibly attributed to the fact that reconstruction occurs on the (113) surface under hydrogen plasma environment which decrease the surface energy.[27] Observations confirm that (113)-grown layers exhibit smooth surface without evident epitaxial features or twins, and photoluminescence (PL) images show limited blue fluorescence indicating that the residual stress remains low. [27] Even high growth rates could produce high quality (113) CVD layers in standard growth conditions with limited number of defects and impurities. Higher partial preferential orientation (73%) of NV centers can be obtained compared with (001) surface with similar coherence times. Therefore, due to the above features, the (113) diamond surface appears as a promising alternative leading to improved NV center fabrication and utilization, [27] which seems to be a good compromise between the readily grown but non-optimal, (001) diamond and ideally oriented although presently technologically not mature (111) diamond.

Beside the photon collection efficiency, the diamond surface should also be suitable for NV quantum sensing because NV centers are implanted within only a few nanometers in depth in order to enhance the spatial range of the sensor for external spins.[33] We call these shallow implanted NV centers briefly "shallow NV centers". The external spins are the objects on top of the diamond surface but the diamond surface itself might also influence the properties of the NV sensor. For instance, the shallow NV centers in the (001) diamond, that is hydrogen terminated after CVD growth, suffer from various problems: it could rapidly convert to neutral state resulting in permanent quenching or temporarily bleaching.[34] It is known for (001) diamond that hydrogen termination leads to a negative electron affinity (NEA) and induces image states at the surface, thus shallow NV centers lose their extra electron and become neutral,[34-36] and surface band bending occurs due to the interaction of this surface with the absorbed water.[37-38] This holds for (111) diamond surface too. Oxygenation of these surfaces converts NEA to large positive electron affinity (PEA) and removes the surface band bending effect. As a consequence, the charge state of shallow NV centers can be preserved in oxygenated (001) and (111) diamonds. However, oxygenated (001) and (111) diamond surfaces induce deep sub-band surface states that shorten the spin coherence time[39] which is detrimental for NV sensing applications.[34, 40-41] Alternatively, through selective $Ag^+$-catalyzed radical substitution of surface carboxyls for fluorine, recent fluorination of nanodiamond could produce a highly hydrophilic interface without etch diamond carbons.[42] However, fluorine is not a bioinert material which would limit the applicability of the diamond NV sensors, thus alternative solutions are needed for bionanosensors. In addition, it brings nuclear spins to the surface that is advantageous for quantum simulations[4] but could be disadvantageous for quantum sensing because it creates an additional noise. It has been recently demonstrated that nitrogenated (001) diamond can be formed in the CVD growth,[43] and accurate theory has shown that this surface possesses PEA and basically no electrical or optical activity[43] that was recently confirmed on direct measurement of the coherence times of shallow NV centers. [44] Furthermore, theory also predicted that nitrogen terminated (111) diamond can be formed by CVD with similar favorable properties like for nitrogen terminated (001) diamond. [36]

According to this summary, the surface termination remarkably influence the properties of the NV sensors, thus deep understanding of the surface terminations is of vital importance to maintain the negative charge state, photostability and coherence time of the NV centers which govern their sensitivity. The effect of surface terminators on (113) diamond on NV properties has not yet been explored, despite the fact, that it is an attractive alternative for realization of NV quantum sensors.

In this work, we employed the first-principles calculations to investigate the geometric and electronic structures of (113) diamond surface with various types of terminators, such as H, O, mixed H/O/OH, F, and N. We show that H- and N-terminated (113) diamond surfaces are not suitable for NV quantum sensors. We also show that oxidation or partial oxidation, forming a H/O/OH mixed termination, is an excellent host of NV quantum sensors due to their positive electron affinity and electrical inactivity, in stark contrast to oxygenated (001) and (111) diamond surfaces. Our simulations indicate that oxygenated (113) diamond surface can be formed with proper choice of precursors and temperature in the CVD process.

## 2. COMPUTATIONAL METHODS

First-principles calculations are performed within the Vienna *ab initio* simulation package (VASP) [45-46] based on density functional theory (DFT) with the projector augmented wave (PAW) method. [47-48] To describe the exchange and correlation interaction, the Perdew-Burke-Ernzerhof (PBE) functional under the generalized gradient approximation (GGA) interaction is used.[49] To get reliable parameters for the surface calculation, we first optimized the primitive diamond bulk. The equilibrium lattice constant of diamond is 3.57 Å under PBE functional. Total energy of bulk shows convergence with $13 \times 13 \times 13$ Monkhorst-Pack[50] scheme Brillouin zone sampling and 400 eV kinetic energy cut-off. A (113)-2×1 reconstruction surface with symmetric tetramer is employed to model (113) surface.[26, 30, 51] Here after thickness convergence test we choose a slab with thirteen bilayers and 15 Å vacuum space to avoid the interaction between periodic slabs. The five bilayers in the middle of the slab are fixed to simulate a bulk-like environment, and the top and bottom four bilayers are fully optimized. The terminator atoms saturate both the top and bottom carbon atoms and eliminate all the dangling bonds so the dipole interaction across the slab can be avoided. A grid of $7 \times 7 \times 1$ k-point sampling was used for slab calculations. The band structure and electron affinity are recalculated with a screened hybrid functional HSE06 of Heyd, Ernzerhof and Scuseria[52-53] with the parameters of 0.2 Å$^{-1}$ for screening and 25% mixing for obtaining accurate results that are directly comparable with experimental data. The geometry structures were optimized until all the force components were smaller than 0.01 eV/Å while the convergence threshold for total energy was set to $10^{-5}$ eV.

## 3. RESULTS AND DISCUSSIONS

Before studying the influence of the surface terminators, it is necessary to briefly discuss the structure of the (113) surface of diamond. The pristine (113) surface contains twofold-coordinated (001) type of facets and threefold-coordinated (111) type of facets with two and one dangling bonds, respectively.[30] The Brillouin zone and high symmetry k-points are shown in Figure 1b. Top and side views of the optimized (113) surface with/without reconstruction are depicted in Figure 1c and d, respectively. After surface reconstruction, the outmost carbon atoms bond with each other, forming tetramer configuration with one carbon dimer, and leaving one dangling bond per atom. The C-C distance of the dimer is 1.53 Å, and the bond length between the higher and lower lying carbon atoms is 1.37 Å. This tetramer reconstruction has been theoretically confirmed to be the most stable reconstruction since a strong strain in the subsurface layers caused by non-optimal bonding of atoms and the asymmetric reconstructions are forbidden. [30] In the following sections, we investigate different terminators that saturate the surface dangling bonds in this tetramer reconstruction model.

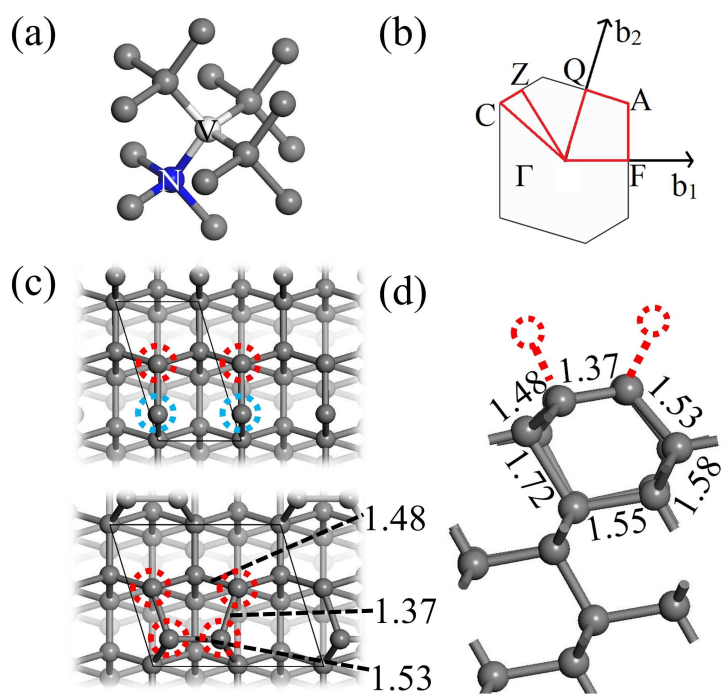

**Figure 1.** (a) Schematic figure of nitrogen-vacancy centre in diamond. The blue and white balls denote the nitrogen and vacancy, respectively. (b) The Brillouin zone of the (113)-2×1 diamond surface. (c) Top view of the (113) surface before (top panel) and after (lower panel) 2×1 reconstruction. The threefold and twofold carbon atoms are denoted by red and cyan dashed circles, and the corresponding bond lengths are also presented in Å unit. (d) The side view of the (113)-2×1 reconstruction surface.

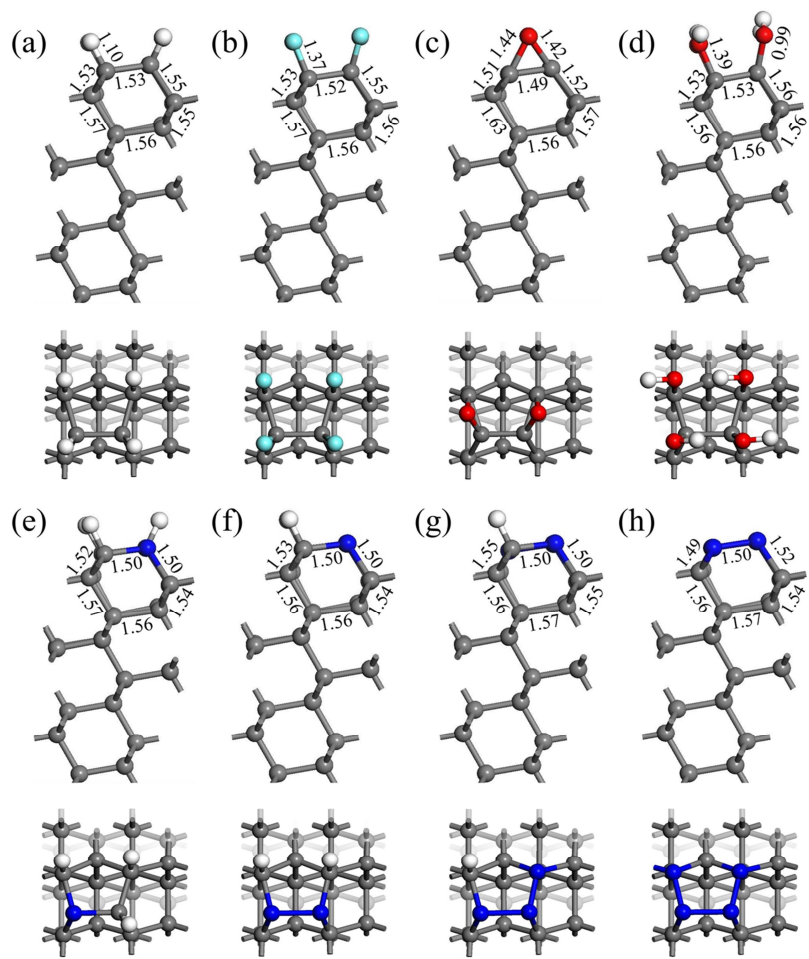

**Figure 2.** Top and side views of the optimized (113)-2×1 diamond surface. (a) H-termination, (b) F-termination, (c) O-termination, (d) OH-termination, (e) N1-termination, (f) N2-termination, (g) N3-termination, (h) N4-termination. The white, green, red, blue, and grey balls represent hydrogen, fluorine, oxygen, nitrogen, and carbon atoms, respectively.

The optimized geometries with different surface terminators are plotted in Figure 2, including H-, F-, O-, OH-, and N-termination. Especially for N-termination, different nitrogen coverage conditions from 1/4 ML to 1ML are considered. These surface termination models are ideal in the sense that they are atomically smooth without any dangling bonds left. The calculated bond lengths near the surface are presented in Figure 2 and are very close to those on (001) diamond surface. [34, 40] Compared to the pristine (113) surface, the original 1.37 Å C-C bonds between the higher and lower carbon atom are stretched to roughly 1.50 Å. For H- and F-termination, four outmost carbon atoms in a 2×1 unit cell would bond with four hydrogen or fluorine atoms, respectively. The relaxed bond length of the higher-lying carbon dimer is 1.63 Å which shows that the H- and F-terminators stretch the dimer outward.

For O-termination, several possible configurations are calculated in order to find the most stable one. To saturate the four dangling bonds two oxygen atoms were bonded to the pristine (113) surface as demonstrated in Fig. 2(c). We found that the ether-like (C-O-C bridge) structure has the lowest energy in which one oxygen atom resides in the bridge between the higher-lying and lower-lying carbon atoms. The C-C distance between the higher-lying and lower-lying carbon atoms is reduced to 1.49 Å due to the "drag" effect of the oxygen atom. Next, the hydroxyl group brings an half spin to the system, therefore similar to hydrogen and fluorine terminations, each 2×1 supercell contain four hydroxyl groups to form a stable closed-shell singlet ground state. We note that the lower energy structure is the one with antiparallel hydroxyl group arrangement, which means the O-H bonds point to opposite directions, rather than parallel configuration which the O-H bonds point to same directions, shown as the top view in Fig. 2(d). [54] The O-H bond formed above the C-C higher-lying dimer is 1.72 Å while it is 1.68 Å at C-C lower-lying dimer.

Let us turn to the N-termination cases. As shown in Fig. 2(e)-(h), we consider different nitrogen coverage conditions using one, two, three, and four nitrogen atoms to replace the outmost carbon atoms, respectively. Furthermore, the remaining carbon atoms on surface without nitrogen substitution are saturated by hydrogen atoms. [43] As expected the nitrogen substitution does not induce significant geometric change. The distance between nitrogen atom and the neighbor carbon atoms is 1.50 Å.

The HSE06 band structure and electron affinity are shown in Figure 3 and Figure S1, respectively. In these band structures, the surface atoms are defined as the terminators and the outmost bilayer carbon atoms. The electron affinity χ is defined as the difference between the vacuum level $E_{vac}$ and the bulk-state conduction band minimum (CBM). In (113)-2×1 H-terminated diamond surface, we find that the H atom could induce deep sub-bandgap states that are surface-related delocalized image states confined above the surface. These image states are also known as Rydberg-states as discussed by Mülliken. [55] Generally the antibonding combinations of the covalent bonds constitute the lowest-energy unoccupied molecular orbitals. However, by increasing the bonding-antibonding energy splitting, high-quantum-number atomic-like state can appear as the lowest-energy unoccupied state of the molecule. This H induced delocalized states also occurred in (001) and (111) diamond surface. [34] H-terminated (113) surface possesses NEA of χ = −1.80 eV. Consequently, the photoexcited electron would occupy these surface unoccupied band and finally get emitted due to the NEA as these surface bands are mixed with the empty in-gap level of NV center. [34] This results in temporary or permanent loss of the excited electron leading to fluorescence intermittency. Thus similar to (001) and (111) H-terminated diamond surface, the H-terminated (113) diamond surface is deficient and detrimental for NV quantum sensors.

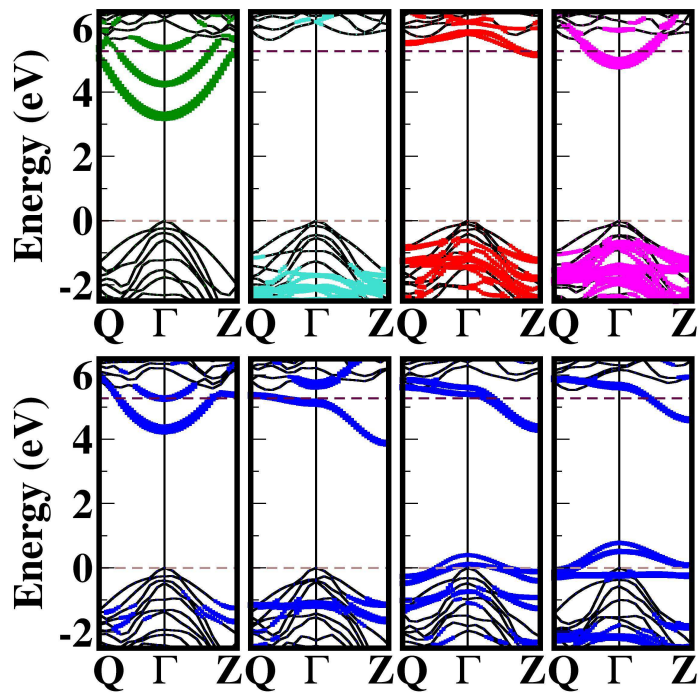

**Figure 3.** The calculated HSE06 band structures of the (113) diamond surface with different terminations. From upper left to right are H-termination, F-termination, O-termination, and OH-termination, respectively; from lower left to right are N1-termination, N2-termination, N3-termination, and N4-termination, respectively. The color curves denote the surface-related bands. The brown dashed line at 0 eV denotes the diamond VBM and the violet dashed line at 5.28 eV represents the diamond CBM. Here we only consider k-lines of Q→Γ→Z because the VBM and CBM lie in this region (see Figure S2).

The fluorinated (113) surface possesses a very large positive electron affinity (PEA) of about +3.30 eV and this value is in good agreement with experiment result on fluorinated (001) surface.[56] Our result shows that for (113) diamond surface the fluorine atoms do not introduce surface-related localized acceptor states below CBM. Thus the fluorescence intermittency or bleaching could be avoided in the photoexcitation process. Using $XeF_2$ precursor for fluorination could avoid the formation of thick Teflon layers on the surface and produce smooth surface. [56] As a consequence, (113) surface with F-termination can be a good host for NV quantum sensor, although, with limited applicability in biological investigations, and the fluorine nuclear spin noise could limit the sensitivity on $^1H$ hydrogen nuclear spins.

Oxygenation is a popular treatment for diamond surface saturation which always yields large PEA and prevents charge conversion of shallow NV centers. Report on (001) surface shows that oxygenation has PEA of +2.4 eV and would introduce deep sub-bandgap occupied bands above the VBM. [34] Additionally, sub-bandgap unoccupied bands also appear below CBM. Previous studies indicate that it is not likely to fully achieve ether-like O-termination, [57] and long-time oxidation would induce surface roughness [35]. Likewise, O-termination of the (111) diamond surface is not a suitable substrate for NV center because of its remarkable reconstructed surface morphology and high chemical activity. [58] However, our results show that these phenomena do not appear in (113) surface. No surface-related bands appear above VBM and the intrusion of empty bands below CBM is small. It has a strikingly favorable PEA of +2.18 eV. These differences may be attributed to the natural conformation of the ether-like bridges at the (113) diamond structures. The stressed bonding configuration of the carbon atom on (001) surface is avoided due to the elimination of continuous chain of C-O-C bridges. [34] The C-O bonds preserve the $sp^3$ bonding character of carbon atom without significant distortion. Since the surface bands reside very close to CBM, the NV center's levels do not interact with these bands. Therefore, O-termination of (113) diamond surface will not influence the single photon absorption in the photoexcitation process of NV center, and direct photoionization does not occur.

Next, we studied other oxygen terminations. OH-termination still provides a slight NEA, $\chi = -0.06$ eV, and the image bands appear at 4.81 eV above VBM. Again the defect states of NV center will mix with these image states that cause permanent loss of luminescence after single photon absorption. Hence the fully hydroxyl (113) surface diamond is not suitable for NV quantum sensor.

Finally, we discuss nitrogen termination. High-quality N-terminated diamond (001) surface can be synthesized by indirect radio frequency $N_2$ plasma process. [43] The predominant component contains substitutional nitrogen and C-H configuration at the surface. Theoretical calculation has already been carried out on (111) diamond surface: N-terminated diamond (111) surface manifests surface electron spin noise free properties which is ideal host for NV centre for quantum simulation of exotic quantum phases and quantum sensing. [36] Nitrogen terminated diamond surface exhibits PEA. These results motivated us to examine nitrogenated (113) surfaces for hosting NV quantum sensors. As shown in Figure 3, single nitrogen substitution (N1) yields unoccupied image bands below CBM due to the C-H bonds. These bands lie at 4.25 eV above VBM. This value is much higher than that in H-termination and lower than that in OH-termination. This N1 configuration yields −0.62 eV NEA and direct photoionization of NV center would occur, therefore N1 configuration is not a good host for NV sensor. NEA surface will convert to PEA surface when the nitrogen coverage reaches 0.5 ML (two nitrogen substitution, denoted as N2). The surface-related image bands lie at 3.91 eV above VBM and the corresponding unoccupied states may mix with the NV centers' empty states. The single photon absorption will cause the photoexcited electron trapped in these image states but not emitted due to the PEA. For three (N3) and four (N4) nitrogen substitution structures, all these surface-related band are shifted upward in the band gap with increasing the nitrogen coverage.

Simultaneously, occupied surface bands appear above VBM. PEA values are improved by increasing the nitrogen coverage. Full coverage (N4) surface yields PEA as +3.56 eV which is larger than that for fluorinated surface. The unoccupied bands appear at 4.33 eV and 4.62 eV above VBM for N3 and N4 configurations, respectively. Therefore N-terminated surface would interfere with the excited state of NV center, and could lead to photoionization in nitrogenated (113) diamond surfaces. We conclude that nitrogen terminated (113) diamond surface is not a favorable host for NV quantum sensor.

One interesting point should be discussed here. The O-terminated (113) surface shows a promising potential utilization for nanosensing technology due to its PEA and electrical inactivity. Remarkably, oxidation is a common and readily accessible treatment of diamond surfaces. Therefore, it is crucial to determine the stability of oxygenated and reconstructed (113)-2×1 diamond surface. To confirm the stability of surface reconstruction of oxygenated (113) diamond, the Gibbs free energies of reconstructed and non-reconstructed (113) surface were calculated. The detailed calculation results can be found in Supporting Information. Our results indicate that the Gibbs free energy of reconstructed 2×1 surface is much lower than non-reconstruction one upon the conditions that the growth temperature is above 500 K and the partial pressure of oxygen molecule under $10^{-8}$ bar (~ $7.5 \times 10^{-6}$ Torr). The relative energy difference favoring the reconstructed surface increases as the temperature increases, indicating the significant thermodynamic stability of reconstructed oxygenated (113) diamond surface at high growth temperature. Previous experiment implemented oxidation of the diamond (001) an (110) surfaces with $2 \times 10^{-6} \sim 5 \times 10^{-5}$ Torr partial pressure of oxygen at various temperature and proposed several kinds of possible surface terminations.[57,59] Atomic oxygen beam treatment generate an oxidized diamond (001) surface without destroying the surface structure and the uptake oxygen was restricted to monolayer.[60] By elegant control of the annealing temperature and oxygen concentration, the reconstructed and oxygenated (113) diamond surface might be achieved via similar methods. [15, 34, 61-62]

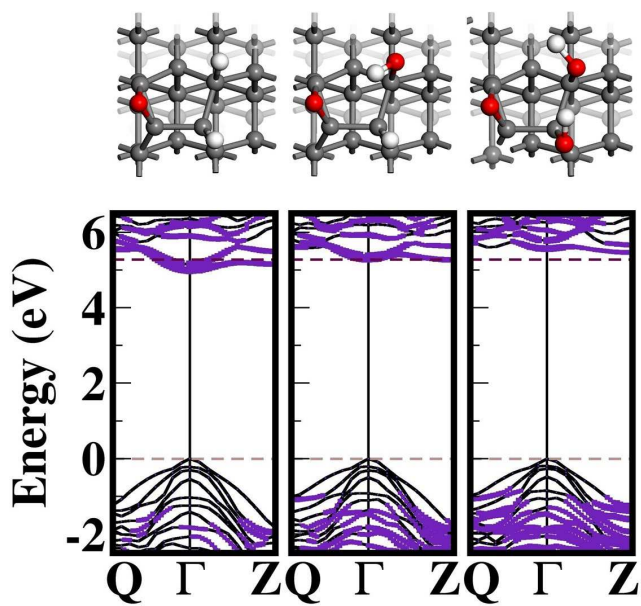

**Figure 4.** The calculated H/O/OH mixed terminations structures and the corresponding HSE06 band structures. The structures, from left to right, are denoted as Mix1, Mix2, and Mix3. The color curves denote the surface-related states. The brown dashed line at 0 eV denotes the diamond VBM and the violet dashed line locating at 5.28 eV represents the diamond CBM.

During the CVD growth of high-quality diamond, it is advantageous to use a high-density plasma at few hundreds of pressures with $H_2/CH_4$ gas mixture and extremely high temperature. [28, 63] On the contrary, PECVD plasmas treatment proceeds under low pressure providing more controllable condition that is useful for the surface functionalization for post-processing, thus a mixture of hydrogen, hydroxyl, and ether-like terminator on the diamond (113) surface may be effectively achieved. Here we consider three kinds of H/O/OH mixed termination of the (113) diamond. Within this 2×1 surface model, each mixed termination contains one C-O-C ether and the remaining dangling bonds are saturated by hydrogen or OH groups, as shown in Figure 4. Similar surface terminators could be formed during the early stage of thermal oxidation of diamond. [35] Compared with the C-O bond length and C-O-C bond angle on oxygenated (001) diamond surface, [34] the C-O bond length and C-O-C bond angle of O-terminated (113) surface decrease and the distortion of the $sp^3$-bonding configuration on the surface carbon atoms is reduced. As a consequence, the strain at the surface induced by oxygenation of (113) diamond is much smaller than that for oxygenated (001) diamond, and results in atomically smooth surface morphology. To a certain extent this may also explain why the fully oxygenated (113) surface possesses deep acceptor state keeping the band gap intact. The results from the mixed termination cases further confirm this observation. Replacing the O bridge with H or OH terminators reduces the steric repulsion between the two oxygen atoms and further pushes up the surface bands below CBM. For the Mix1, Mix2, and Mix3 terminations, the unoccupied surface bands lie at 4.99 eV, 5.27 eV, and 5.46 eV above VBM. The calculated electron affinity are +0.30 eV, +0.47 eV, and +0.87 eV, respectively. As a consequence, increasing hydrogen concentration will induce deeper surface states and lower electron affinity, while increasing OH group concentration will pushes up the position of the surface bands and electron affinity. Surprisingly, due to the existence of C-O-C ether, the NEA of pure H- or OH-terminations are converted to PEA. Such an electron affinity has been measured experimentally on oxygenated diamond surfaces. [35] Obviously, no image states appear in the band gap. Hence, these H/O/OH mixed termination could be optimal candidate for NV quantum sensors.

Here, we would discuss the chemical stability of the surface radicals of O-terminated diamond surface. On this O-terminated (113) diamond surface, the bonding of the C-O-C is characterized by a shorter 1.49 Å C–C bond (backbond C–C is 1.54 Å), two 1.42 C–O bonds, and C–O–C angle of 62.9°, which forms a ring-like structure, known as epoxide in chemistry. This ring-like structure is presumed to be a relatively reactive functional group due to the structural stress. In organic molecules, the epoxide ring can react with water molecule in presence of acid or base catalyst and then the ring opens.[64] However, previous experimental and theoretical works indicated the stable product of epoxide group existing on solid surface, e.g. diamond/water interface [65], diamond (100) [57, 66-67], quartz crystal [68], Si(111) [69], and graphene [70]. Thus, we calculated the reaction energy barriers of water opening epoxide ring procedure and the barriers are in the range of 0.9~1.0 eV, as shown in Fig. S6-S8. The existing methods for epoxidation include ultraviolet radiation,[71-72] electron beam irradiation,[73] and especially high temperature plasma treatment,[57] organic dioxiranes,[66] and ozonolysis of diamond.[67] Hence the epoxide surface with careful treatment may thermodynamically stable and it can be expected to form on (113) diamond surface. Our results indicate the oxygenated (113) diamond surface would potentially be applied on NV quantum sensing; further experimental and theoretical efforts for surface functionalization of diamond are needed towards this direction.

Next, we further investigate the photon collection efficiency of (113) diamond surface since it can lead more reliable and sensitive sensors. Figure S9 demonstrates the structural details of the four possible NV defect orientations of (001), (011), (111), (112), and (113) five common facets of CVD diamonds, and the angles are summarized in Table 1. In (001) surface, α = 54.8° for the four orientations. Experimental result revealed that the photon collection was low with such a large angle of 54.8 degrees between the NV centers' symmetry axis and the normal vector of the

(100) diamond surface. [15, 17] For (111) surface, zero degree angle between the NV center's symmetry axis and the normal vector of the diamond (111) surface lead to maximum photon collection efficiency from NV centers [17, 19, 74], where ~ 100% preferential orientation of NV centers can be achieved for NV centers grown into diamond during well-controlled CVD process. Consequently, for diamond (112) and (113) higher-index surfaces, [29] which possess 19.5° and 29.5° angles w.r.t. <111> axis, a higher photon collection efficiency than that for (001) surface can be expected. We note here that previous reports indicate that the NV defect prefers to grow along the growth direction. [15, 27] Therefore we speculate that the (112) and (113) diamond surfaces may achieve better alignment than (001) surface due to the smaller angles. This good alignment can further favor this host for NV sensing applications. In particular, the preferential alignment of NV centers can be very important in magnetic field camera and related sensor applications, where homogeneous electron spin resonance signals can be observed from ensemble of NV centers. We also list the CVD growth rate of these five common diamond surfaces in Table 1. The (113) diamond surface can achieve higher growth rate than (011) and (111) surfaces which is favorable for fabricating NV center. The ease of growth and processing on (113) diamond surface provide an effective and economical choice to fabricate NV center for sensing application.

**Table 1.** The angles between four orientations of conventional diamond bulk ($<1\bar{1}1>$, $<\bar{1}11>$, $<11\bar{1}>$ and $<111>$) and surface normal of (001), (011), (111), (112), and (113) diamond surfaces. The growth rate (μm/h) is also listed.

|  | (001) | (011) | (111) | (112) | (113) |
|---|---|---|---|---|---|
| $<1\bar{1}1>$ | 54.8° | 90.0° | 109.5° | 61.9° | 58.5° |
| <$\bar{1}11$> | 54.8° | 35.2° | 109.5° | 61.9° | 58.5° |
| <$11\bar{1}$> | 54.8° | 90.0° | 109.5° | 90.0° | 100° |
| <111> | 54.8° | 35.2° | 0.0° | 19.5° | 29.5° |
| Growth rate | $5^{75}$-$150^{63}$ | >$10^{76}$ | 6-$12^{62}$ | NA | 15-$50^{27}$ |

## 4. CONCLUSIONS

In summary, the effects of H, O, OH, O/H/OH mixed, F, and N terminators of (113) diamond surface on the NV sensors, are investigated by means of first principles calculations. Hydrogen and hydroxyl terminated (113) diamond surface induces NEA and sub-bandgap image states which would mix with NV center's excited state, leading blinking or quenching of the luminescence. Fluorine terminated (113) surface can be a feasible host for NV quantum sensor due to its large PEA but with limited resolution of $^1$H nuclear spins. Nitrogen termination could also transform from NEA to PEA when the coverage is larger than 0.5 ML. The image bands will be pushed up in the band gap and converted to localized unoccupied surface states by increasing the nitrogen coverage. Nitrogen terminated (113) diamond surface will results in blinking in the luminescence of NV center. Interestingly, unlike (001) and (111) diamond surfaces, oxygenated (113) diamond surface does not introduce bands that can directly interact with the NV center, thus fluorescence intermittency or bleaching can be avoided in the single photon absorption process. The stability of reconstructed and oxygenated (113)-2×1 diamond surface is investigated by calculating the Gibbs free energy. The simulations imply high stability of these surfaces during CVD growth of diamond with conventional CVD growth conditions. Our results also reveal that fully oxygenated diamond (113) surfaces are not required and mixed H/O/OH terminators are also potential candidates, which may be created during the early stage of thermal oxidation. Because of these impressive advantages, the oxygenated (113) diamond surface, as a technologically mature material, is a very prospective host for NV quantum sensing applications.

**ASSOCIATED CONTENT**

**Supporting Information:**

The HSE06 calculated electron affinity, PBE band structures, and calculated angles between four orientation and magnetic field for (001), (011), (111), (112) and (113) surfaces are shown. Discussions of detailed Gibbs free energies calculations and chemical stability of O-terminated (113) surface are also included.

**AUTHOR INFORMATION**

**Corresponding authors**

(J. C) Email: jpchou@cityu.edu.hk

(A. H) Email: alicehu@cityu.edu.hk

**ACKNOWLEDGMENT**

S. L., J.-P. C., and A. H. acknowledge the funding support from City University of Hong Kong under the project 9610336. A. G. acknowledges the support from the National Research Development and Innovation Office of Hungary (NKFIH) within the Quantum Technology National Excellence Program (project no. 2017-1.2.1-NKP-2017-00001), EU QuantERA Q-Magine (NKFIH Grant no. 127889), and EU H2020 ASTERIQS project.

# Supporting Information:
# Oxygenated (113) diamond surface for nitrogen-vacancy quantum sensors with preferential alignment and long coherence time from first principles

*Song Li[1], Jyh-Pin Chou[1,\*], Jie Wei[1], Minglei Sun[2], Alice Hu[1,\*], and Adam Gali[3,4]*

[1] Department of Mechanical Engineering, City University of Hong Kong, Kowloon Tong, Hong Kong, China

[2] School of Mechanical Engineering, Southeast University, Nanjing, Jiangsu 211189, China

[3] Institute for Solid State Physics and Optics, Wigner Research Centre for Physics, Hungarian Academy of Sciences, Budapest, POB 49, H-1525, Hungary

[4] Department of Atomic Physics, Budapest University of Technology and Economics, Budafoki út 8, H-1111, Budapest, Hungary

Corresponding Authors: (J. C.) Email: jpchou@cityu.edu.hk
(A. H.) Email: alicehu@cityu.edu.hk

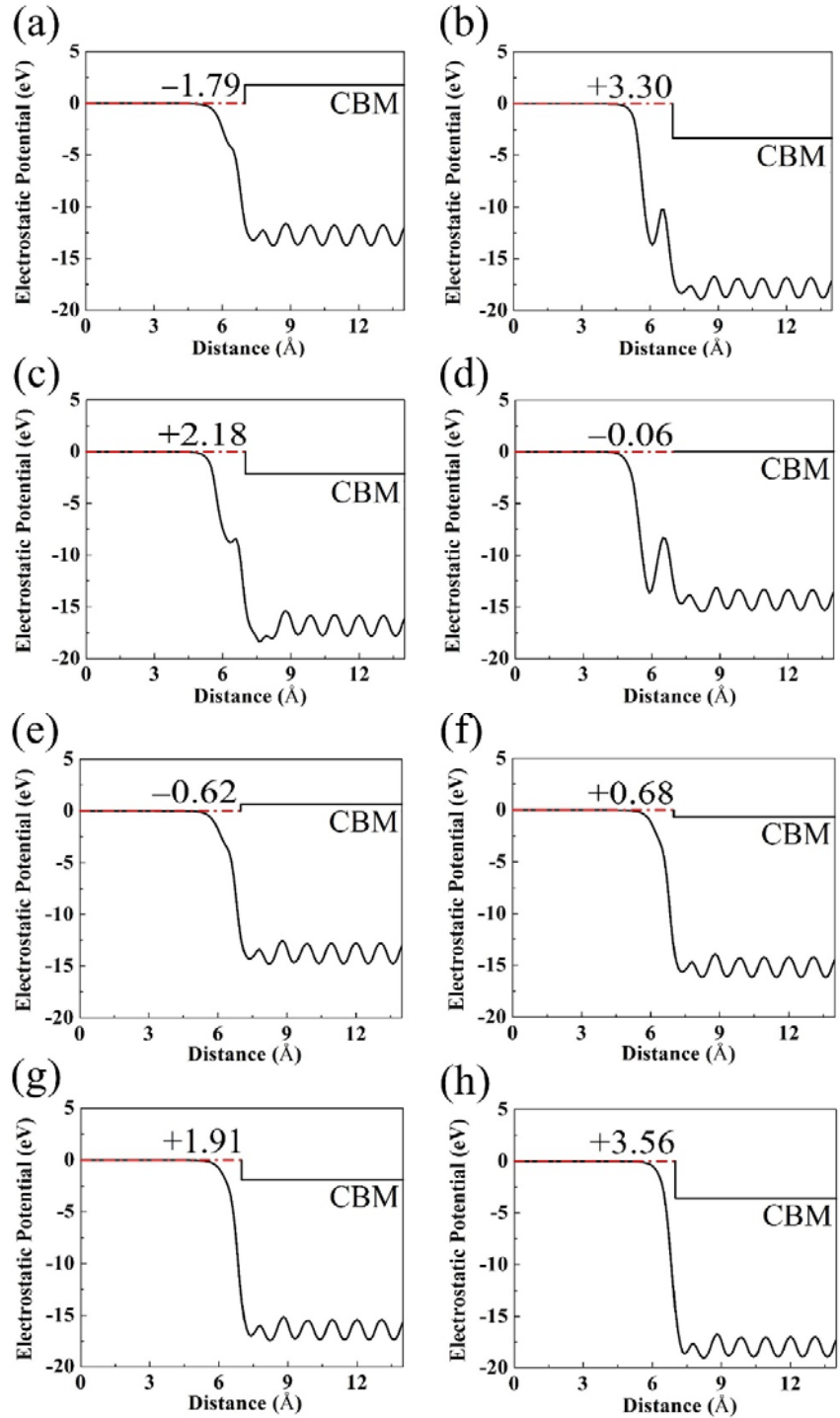

**Figure S1.** The HSE06 calculated electron affinity of the (113)-2×1 diamond surface with various different terminations. (a) H-termination, (b) F-termination, (c) O-termination, (d) OH-termination, (e) N1-termination, (f) N2-termination, (g) N3-termination, (h) N4-termination. The calculated average potential vs. the $z$ coordinate is demonstrated, where the $z$ coordinate is the direction perpendicular to the slab. The red dash-dot lines denote the vacuum level.

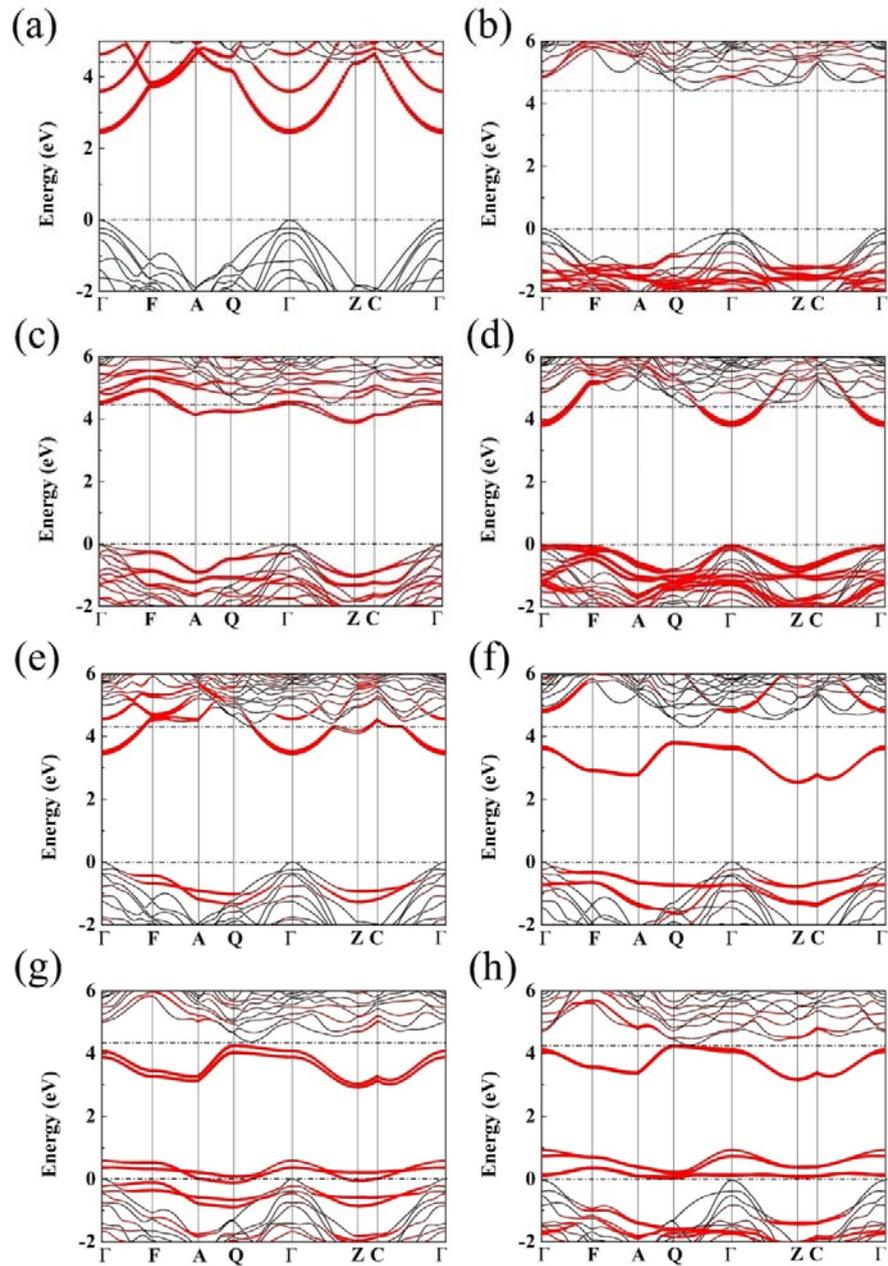

**Figure S2.** The calculated PBE band structure of the (113) diamond surface with various different terminations. (a) H-termination, (b) F-termination, (c) O-termination, (d) OH-termination, (e) N1-termination, (f) N2-termination, (g) N3-termination, (h) N4-termination. The red line denotes the surface-related bands.

## Stability of O-terminated (113)-2×1 surface

The formation energies of O-terminated non-reconstructed and reconstructed (113) surface are calculated by using two-side surface model in 2×1 unit cell. To saturate all the dangling bonds, there are three oxygen atoms on non-reconstructed and two oxygen atoms on reconstructed surfaces. The possible configurations and their energy difference are shown in Figure S3 that the most stable configurations are set as the reference points. The formation energy in zero temperature is calculated by [1-2]:

$$E_{\text{formation}} = [E_{\text{tot}} - N(O)E(O) - N(C)E(C)]/N_{\text{tot}}$$

The $E_{\text{tot}}$ represents the total energy of reconstructed or non-reconstructed structures. The $N(O)$, $N(C)$, and $N_{\text{tot}}$ represent the number of oxygen atoms, the number of carbon atoms, and total number of atoms, respectively. The number of oxygen atoms is four and six for reconstruction and non-reconstruction surfaces, respectively. The $E(O)$ is half of the energy of one oxygen molecule and $E(C)$ is the energy per atom of diamond bulk.

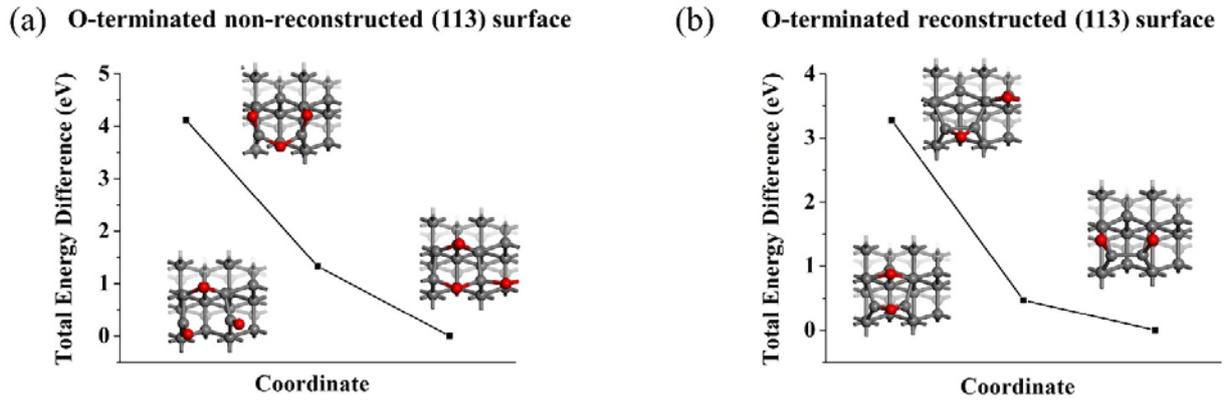

**Figure S3.** The total energies differences of possible configurations of (a) O-terminated non-reconstructed (113)-2×1 surface and (b) O-terminated reconstructed (113)-2×1 surface. The most stable configuration of reconstructed and non-reconstructed (113) surface are obtained.

To determine the relative stabilities of O-termination on (113)-2×1 diamond, the Gibbs free energies [1-2] are introduced since the oxygen content would vary under different experimental conditions:

$$G = E_{\text{formation}} - N(O)\mu(O)$$

where $\mu(O)$ is the chemical potential of oxygen atom which can be determined through [1-2],

$$\mu(O) = H°(T) - H°(T_R) - TS°(T) + k_B T\left(\frac{P}{P_0}\right).$$

The $\mu(O)$ is a function of temperature $T$ and oxygen partial pressure $P$. $H^o$ and $S^o$ are the enthalpy and entropy of the oxygen atom, respectively, where $P_0$ = 1 bar and $T_R$ = 298.15 K. The temperature and pressure dependence of enthalpy and entropy can be obtained from the CRC handbook [3]. The detailed results are listed in Table S1. The formation energy difference per atom is $\Delta E_{formation}$. To ensure the convergence of the calculations, we enlarge the thickness of models from 13 bilayers to 37 bilayers.

**Table S1**. The calculated energies for each component are given. All the units are in eV.

|  | PBE(13 layers) | PBE(37 layers) |
|---|---|---|
| $E_{re}$ | −486.85050 | −1360.1963 |
| $E_{non}$ | −499.06080 | −1372.4144 |
| $E(C)$ | −9.10007 | −9.10007 |
| $E(O)$ | −4.92364 | −4.92364 |
| $E_{formation}(re)$ | 6.048 | 6.309 |
| $E_{formation}(non)$ | 3.685 | 3.937 |
| $\Delta E_{formation}$ | 0.044 | 0.016 |

For the two most stable reconstructed and non-reconstructed structures, the Gibbs free energies as a function of $\mu(O)$ are plotted in Figure S4 (a). It clearly shows the reconstructed surface has lower Gibbs free energy than the non-reconstructed surface when $\mu(O) < -1.185$ eV. To provide a further insight into this condition, Figure S4 (b) demonstrates the Gibbs free energy as a function of pressure and temperature. For partial pressure in the range from $10^{-10}$ bar to $10^0$ bar, the non-reconstructed surface is more stable than the reconstructed one under the temperature condition of 700 K. For common CVD deposition of diamond, the temperature is about 900 ~ 1500K [4-7] which indicates that the diamond (113)-2×1 surface prefers to form reconstructed O-termination. Previous experiment realized oxidation of the diamond (100) surface with $2 \times 10^{-6}$ ~ $5 \times 10^{-5}$ Torr partial pressure of oxygen at 1100 ~ 1300 °C [8]. Hence, the reconstructed and oxygenated (113) diamond surface can be easily accomplished via controlling the annealing temperature and oxygen concentration.

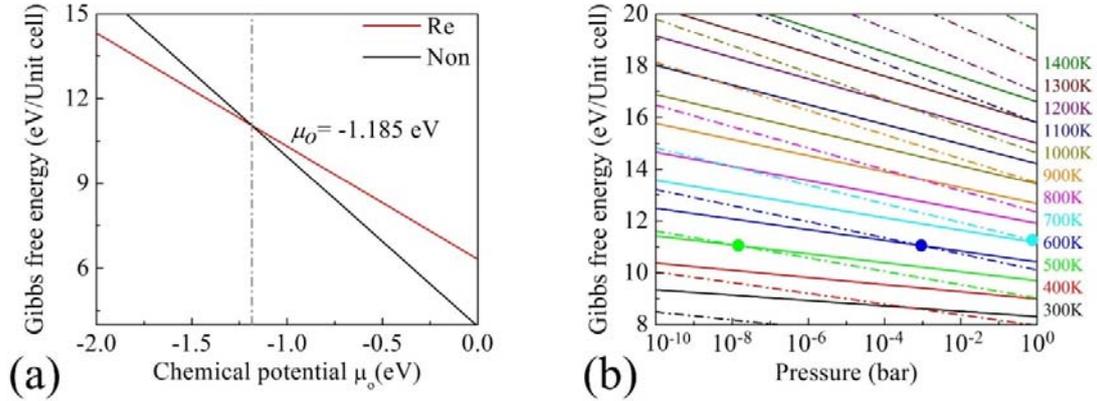

**Figure S4.** (a) The calculated Gibbs free energy as a function of oxygen chemical potential $\mu(O)$. The red and black lines represent the results of reconstruction and non-reconstruction surface, respectively. The transition occurs at $\mu(O) = -1.185$ eV. (b) The calculated Gibbs free energy versus partial pressure of oxygen under different temperature. The solid (dash) line represents the reconstructed (non-reconstructed) oxidization surface. The solid dots denote the transformation point at temperature of 500K (green), 600K (deep blue), and 700K (light blue).

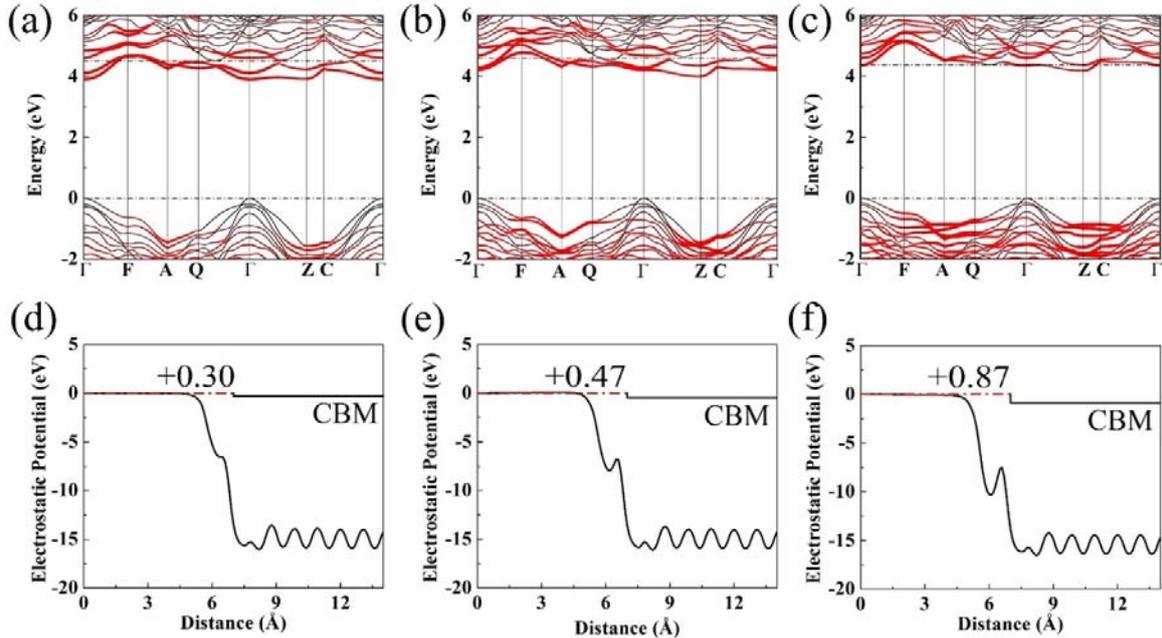

**Figure S5.** The calculated PBE band structure and HSE electron affinity of the H/O/OH mixed terminations on (113) diamond surface. (a) (d): Mix1; (b) (e): Mix2; (c) (f): Mix3 terminations.

# Chemical Stability of O-terminated (113)-2×1 surface

Because limited published literatures on oxidized diamond surface, we tried to provide more useful information by gathering investigations on solid surface, such as graphene, a counterpart of diamond. Research on graphene oxide (GO) may offer reasonable evidence that can support the chemical stability of epoxide structure on diamond surface to a certain extent. Considerable experimental characterizations including solid-state nuclear magnetic resonance indicated the existence of epoxide on graphene [9-14]. Considering the chemical stability of epoxide on graphene, previous investigation used sodium ethoxide (NaOEt), one kind of strong nucleophile, to attack GO [9]. The nucleophilic attack was slow and only part of the epoxide rings would be opened. Moreover, the previous researches [9, 15] indicated that the oxygen layers on GO basal plane could prevent nucleophilic attack on carbon atoms, providing reliable reason for the chemical inactivity of the epoxide structures on GO. These experiment observations imply that the epoxide on diamond surface is unlikely to be opened by weak nucleophile at ambient conditions.

Previous theoretical calculations reveal water and epoxide group formation from OH groups was exothermic by 0.45 eV and the energy barrier was 0.41 eV [11, 14]. Correspondingly, to generate OH group through reaction between water and epoxide, the energy barrier was about 0.9 eV, indicating that water attack epoxide group is unlikely occur at room temperature. Regarding diamond surface, the reaction energy barriers for water attacking epoxide group were calculated here by the Nudged Elastic Band (NEB) method [16-17]. We considered two migration ways of water molecule to attack epoxide group as shown in Fig. S6 and S7. To open the ring of epoxide, the reaction energy barriers are 0.99 eV and 0.88 eV for the two paths, respectively (see Fig. S8). Therefore, we speculate at least in ambient condition it is difficult to open the ring of epoxide for water.

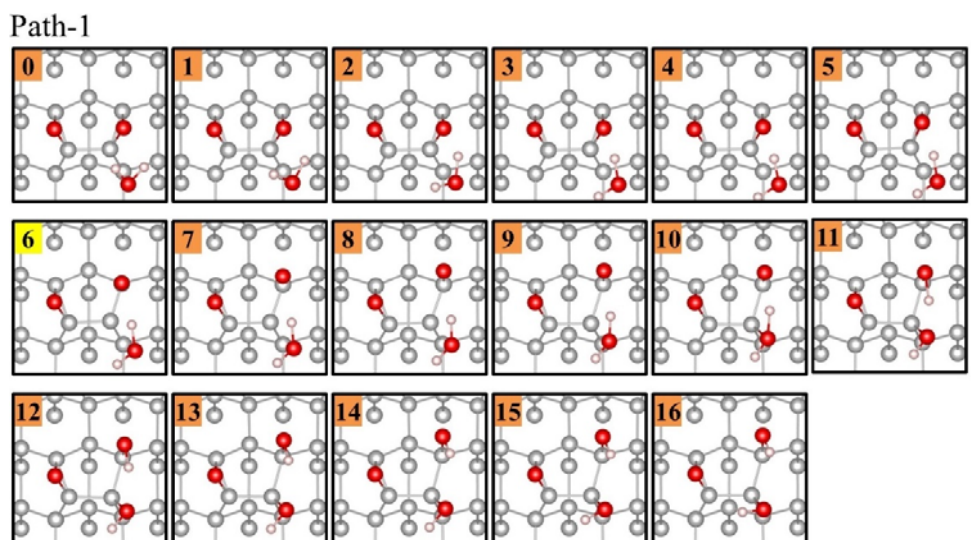

**Figure S6.** The structures of NEB calculation of path-1.

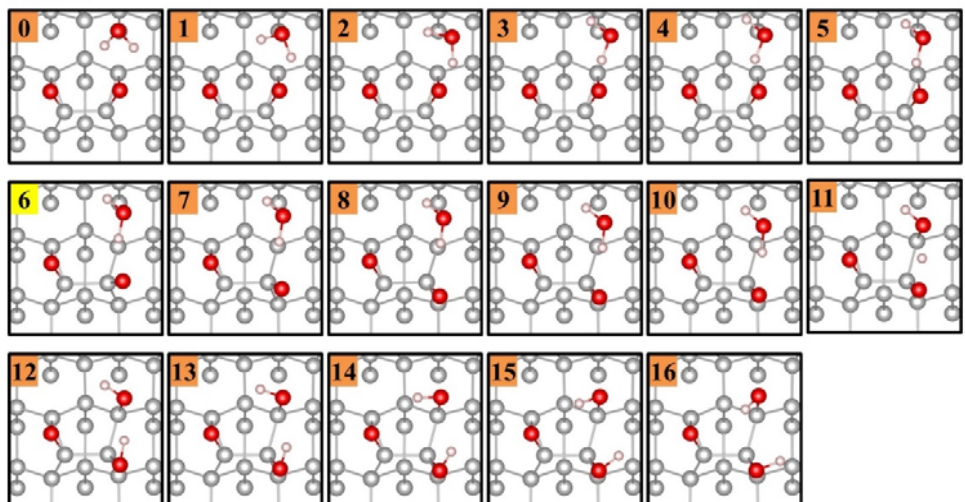

**Figure S7.** The structures of NEB calculation of path-2.

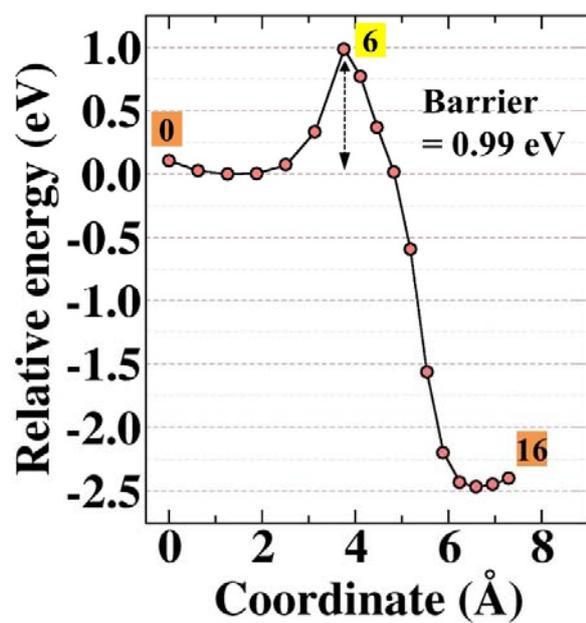
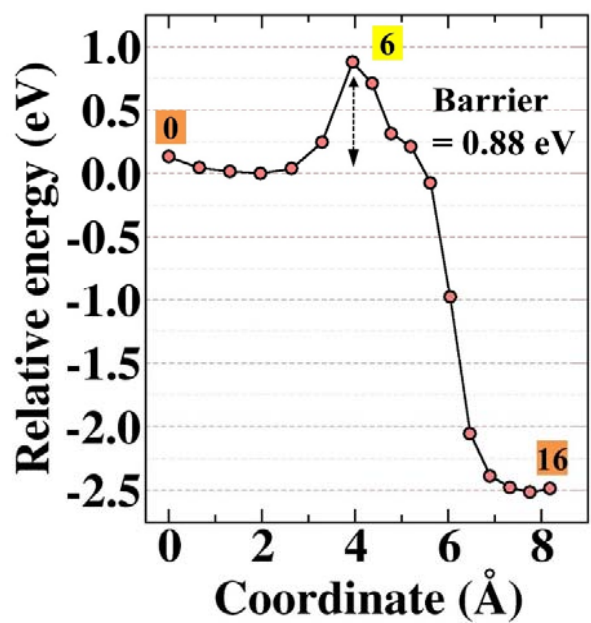

**Figure S8.** The calculated reaction energy barriers of two attacking paths.

**Photon collection efficiency**

High efficient photon collection in NV center is indispensable for extensive application in sensitive sensing and quantum information processing. The optimal collection efficiency for NV center in nanopillars could be achieved when the emitting dipole is oriented perpendicularly to the pillar's axis, which can be realized on same aligned NV center and nanopillar along (111) axis.[18-19] The selection of surface facet becomes an important issue, as shown in Fig. S9. The crystal characteristics result in four possible crystallographic axes of NV centers: <1$\bar{1}$1>, <$\bar{1}$11>, <11$\bar{1}$> and <111>. For our (113) surface, a remarkable preferential orientation as high as 73% of NV center along <111> was observed.[20] We suspect that this small angle deviation from growth direction with high preferential orientation may achieve desirable photon collection efficiency.

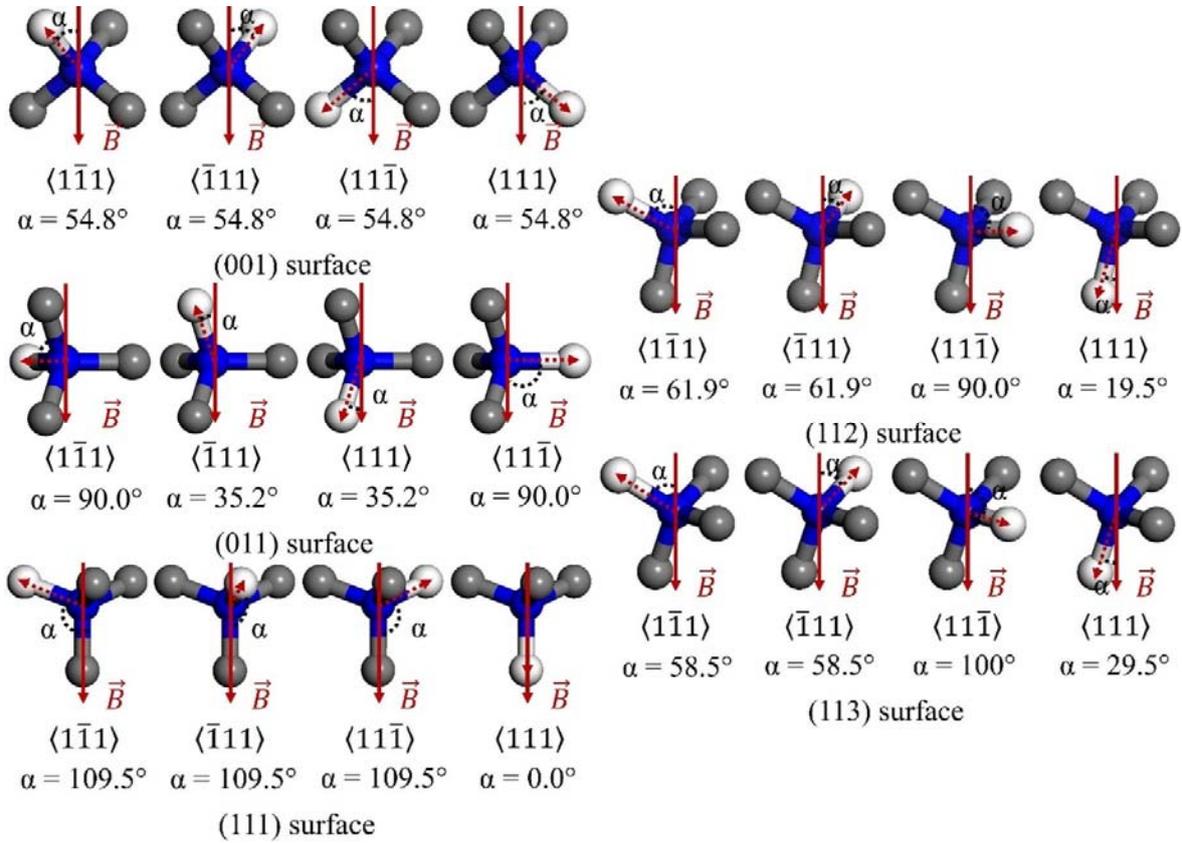

**Figure S9.** The calculated angles between four orientation (<1$\bar{1}$1>, <$\bar{1}$11>, <11$\bar{1}$> and <111>) and magnetic field for (001), (011), (111), (112) and (113) surfaces. The white, blue and grey balls denote the vacancy, nitrogen atom, and carbon atom, respectively. The red line denotes the direction of magnetic field.